\documentclass[reprint,showpacs,twocolumn,superscriptaddress,aps,prl]{revtex4-2}
\usepackage[utf8]{inputenc}
\usepackage[american,british]{babel}
\usepackage[T1]{fontenc}
\usepackage[pdftex]{graphicx}  
\usepackage{graphicx, xcolor}
\usepackage{dcolumn}
\usepackage{braket}
\usepackage{bm}
\usepackage{amsmath,amsthm,amssymb}
\usepackage{color}
\usepackage{verbatim}
\usepackage{ulem}

\definecolor{darkGreen}{RGB}{0,110,0}
\definecolor{darkBlue}{RGB}{0,0,130}
\usepackage{hyperref}

\hypersetup{
 colorlinks=true,
 linkcolor=blue,
 anchorcolor = blue,
 citecolor = blue,
 filecolor = blue,
 urlcolor = blue
}

\usepackage{dsfont}
\usepackage{physics}

\begin{document}
\title{An entanglement asymmetry study of black hole radiation}
\date{\today}

 \author{Filiberto Ares}
 \affiliation{SISSA and INFN, via Bonomea 265, 34136 Trieste, Italy}

  \author{Sara Murciano}
 \affiliation{Walter Burke Institute for Theoretical Physics, Department of Physics and IQIM, Caltech,
    1200 E. California Bl., Pasadena, CA 91125, USA}

 \author{Lorenzo Piroli}
 \affiliation{Dipartimento di Fisica e Astronomia, Università di Bologna and INFN,
Sezione di Bologna, via Irnerio 46, I-40126 Bologna, Italy}

 \author{Pasquale Calabrese}
 \affiliation{SISSA and INFN, via Bonomea 265, 34136 Trieste, Italy}
 \affiliation{The Abdus Salam International Center for Theoretical Physics, Strada Costiera 11, 34151 Trieste, Italy}

\begin{abstract}

Hawking's discovery that black holes can evaporate through radiation emission has posed a number of questions that with time became fundamental hallmarks for a quantum theory of gravity.
The most famous one is likely the information paradox, which  finds an elegant explanation in the Page argument suggesting that a black hole and its radiation can be effectively represented by a random state of qubits. 
Leveraging the same assumption, we ponder the extent to which a black hole may display emergent symmetries, employing the entanglement asymmetry as a modern, information-based indicator of symmetry breaking. 
We find that for a random state devoid of any symmetry, a  $U(1)$ symmetry emerges and it is exact in the thermodynamic limit before the Page time. 
At the Page time, the entanglement asymmetry shows a finite jump to a large value.
Our findings imply that the emitted radiation is symmetric up to the Page time and then undergoes a sharp transition.
Conversely the black hole is symmetric only after the Page time.

\end{abstract} 

\maketitle
\paragraph{\it Introduction ---} 
Black holes usually put quantum mechanics on the spot. A long
standing puzzle is whether information is lost after they evaporate,
contradicting the unitarity of quantum mechanics~\cite{hawking-75,hawking-76}. 
In the nineties, Don Page took a significant stride towards its understanding with a very simple argument~\cite{page-93, page-93-2}: 
by approximating the black hole and the radiation with a random state of qubits, he showed that  
quantum unitarity implies that the entanglement entropy of the radiation emitted in the
process first grows and then it decreases, going back to zero when the black hole disappears.
A series of recent calculations~\cite{aemm-19, penington-20, ahmst-20, pssz-22} have shown 
in specific quantum gravity models using replica wormholes that the entanglement entropy of the
radiation actually follows this behaviour, and information is preserved.

A natural follow up question is: what are the implications of quantum unitarity to 
symmetries in the evaporation of a black hole? Many different analysis
have argued that exact global symmetries should be explicitly broken in consistent quantum gravity models~\cite{mw-57, klls-95, bd-88, bs-11, ho-19, ho-21}. In particular, a unitary black hole evaporation would forbid the existence of global symmetries~\cite{hs-21}. Motivated by these 
works, we examine here how a broken global $U(1)$ symmetry evolves during the black hole evaporation using the original model of Page, 
which assumes that the dynamics of the black hole is highly chaotic.
A very suitable tool for this purpose is the entanglement asymmetry, 
a quantity that measures how much a symmetry is broken in a part of a 
system, which has been recently introduced in Ref.~\cite{amc-23} in the context 
of non-equilibrium many body quantum systems, see also~\cite{fac-23b,amvc-23,cm-23,rka-23,makc-23,cv-23,cc-23, bkccr-23}.

\paragraph{\textit{Basic setup ---}}
Let us consider a quantum system consisting of two spatial parts $S=A\cup B$ in which the Hilbert space factorises as $\mathcal{H}_A\otimes \mathcal{H}_B$, being $d_A$ and $d_B$ the dimensions of $\mathcal{H}_A$ and $\mathcal{H}_B$, respectively. In the analysis of
the evaporation of a black hole, one of the subsystems, e.g. $B$, represents the black hole while the complementary, $A$, is identified with the radiation emitted during the process. 
Initially, the subsystem corresponding to the radiation is empty, $A=\varnothing$, and, therefore, $B=S$. As the black hole evaporates, the degrees of freedom of $B$ are transferred to $A$ until $B=\varnothing$ and $A=S$, point at which the black hole is completely evaporated. 
Accordingly in this process the time can be identified as $t\propto \log d_A$.
A usual assumption to characterize this process  is to presume that the state shared by $A$ and $B$ at any time is a typical pure state~\cite{page-93-2}. The crucial result of Page~\cite{page-93} is that, if we take a random ensemble of pure states distributed according to the Haar measure and we average their entanglement entropies $S_1$, we find that, up to subleading corrections,
\begin{equation}\label{eq:page_curve}
\mathbb{E}[S_1]=\min(\log d_A,\log d_B)+O(1),
\end{equation}
where $\mathbb{E}[\cdot]$ denotes the average over the ensemble of Haar random states.
%
%
Eq.~\eqref{eq:page_curve}, commonly referred to as the Page curve, indicates that the entanglement entropy of the emitted radiation (and of the black hole) initially grows at maximum rate; this is attributed to the creation of entangled pairs between the degrees of freedom trapped inside the black hole and those escaping to the region $A$ ($d_A<d_B$). 
The entanglement reaches a maximum when $d_A=d_B$, defining the Page time.
As the black hole continues to evaporate beyond the Page time, the (entanglement) entropy decreases. This reduction is physically explained by the emission of particles already entangled with the radiation.
Ultimately, the entanglement entropy reaches zero when the black hole is entirely evaporated ($\log d_B = 0$). This implies that there is no loss of information, thereby reconciling black hole evaporation with unitary quantum evolution.

In this paper, we consider generic Haar random pure states breaking any possible symmetry. 
Our focus is on understanding the destiny of a broken global $U(1)$ symmetry by assessing the entanglement asymmetry of $A$, that quantifies the extent to which symmetry is broken within the subsystem.
In the context of an evaporating black hole, our calculation corresponds to the time evolution of the asymmetry of the radiation. 
The same problem was also addressed in Ref.~\cite{lntt-22} but without the lens of entanglement asymmetry.

For simplicity, we choose a system of $L=\ell_A+\ell_B$ qubits, i.e. $d_A=2^{\ell_{A}}$ and $d_B=2^{\ell_B}$.
The local basis for each qubit $k=1,\dots, L$ is given by the states $|0\rangle_k$ and $|1\rangle_k$, with global vacuum $|0\rangle=\otimes_k|0\rangle_k$. 
We prepare the system in a pure state $U\ket{0}$, where $U$ is a $2^L\times 2^L$ unitary matrix taken from the Haar random ensemble. 
The state of the subsystem $A$
is described by the reduced density matrix $\rho_A=\Tr_{B}(U\ket{0}\bra{0}U^\dagger)$. We further consider the charge operator $Q=\sum_{k=1}^L\ket{1}_k\bra{1}_k$, which counts the number of excitations and generates a global $U(1)$ symmetry group. This charge is local in the sense that $Q=Q_A+Q_B$. 
A generic matrix $U$ provides a state $U\ket{0}$ that is not an eigenstate of $Q$. As a consequence, $[\rho_A, Q_A]\neq 0$ and, therefore, $\rho_A$ breaks the $U(1)$ symmetry generated by $Q$.
This is different from considering states that preserve the symmetry $Q$ and leading to symmetry resolved Page curves~\cite{bd-19,lntt-22,mcp-22,ll-23}.

In this setup, the entanglement asymmetry~\cite{amc-23} is defined
by introducing the auxiliary reduced density matrix $\rho_{A, Q}=\sum_q\Pi_q \rho_A \Pi_q$, with $\Pi_q$ the projector on the eigenspace of $Q_A$ of charge $q \in \mathbb{Z}$, satisfying $[\rho_{A, Q}, Q_A]=0$. 
We can then define the entanglement asymmetry in terms of the $n$-R\'enyi entropy $S_n(\rho)=\frac1{1-n}\log\Tr(\rho^n)$ as 
\begin{equation}\label{eq:def_ent_asymm}
\Delta S_A^{(n)}=S_n(\rho_{A, Q})-S_n(\rho_A).
\end{equation}
The von Neumann asymmetry is obtained in the limit $n\to1$.
We will show that the choice of $n$ is not important for the resulting physics.  
The entanglement asymmetry satisfies two essential
properties to be a quantifier of symmetry breaking~\cite{hms-23}: it is non-negative $\Delta S_A^{(n)}\geq 0$ and $\Delta S_A^{(n)}=0$ if and only if $[\rho_{A}, Q_A]=0$; that is, when $\rho_A$ respects the symmetry generated by $Q$ (i.e. $\rho_{A,Q}=\rho_{A}$).
In the context of resource theory, $\Delta S_A^{(1)}$ has been also studied in Refs. \cite{r1,r2,r3,r4}.

\textit{Analytic calculation for integer $n\geq 2$ ---} 
To calculate analytically the entanglement asymmetry over an ensemble of Haar random states $\{U\ket{0}\}$,  we assume that we may replace $\mathbb{E}[\log \mathrm{Tr}(\rho_{A,Q}^n)]$ with $\log \mathbb{E}[\mathrm{Tr}(\rho_{A,Q}^n)]$,  (same for $\rho_A$). Although the two expressions are not strictly equal, we check numerically that bringing the average “inside the logarithm” is a good approximation up to exponentially small sub-leading corrections in the system size $L$. This may be expected based on arguments exploiting the concentration of Haar measure.

\begin{figure}
     \centering
     \includegraphics[width=\linewidth]{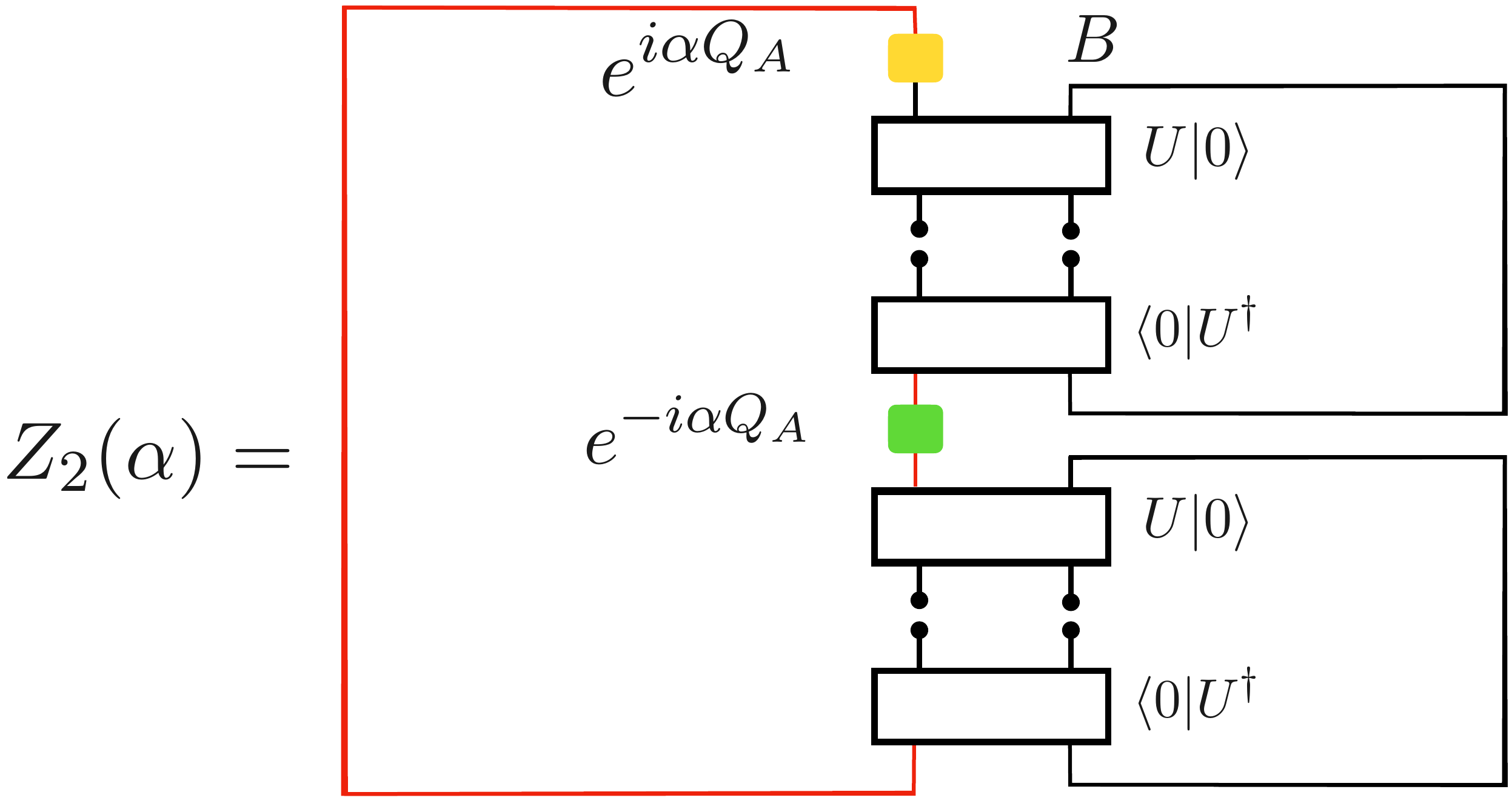}
     \caption{Graphical representation of the charged moments $Z_2(\alpha)$  in Eq. \eqref{eq:def_charged_mom}.}
     \label{fig:circuit}
\end{figure}

The main advantage of this approximation is that, using the Fourier representation of the projector $\Pi_q$, $\mathbb{E}[\Tr(\rho_{A, Q}^n)]$ can be written as
\begin{equation}\label{eq:renyi_rhoAQ}
\mathbb{E}[\Tr(\rho_{A, Q}^n)]=\int_{-\pi}^\pi\frac{d\alpha_1\cdots d\alpha_n}{(2\pi)^n}
\mathbb{E}[Z_n(\boldsymbol{\alpha})],
\end{equation}
in terms of the averaged charged moments of $\rho_A$~\cite{amc-23}
\begin{equation}\label{eq:def_charged_mom}
Z_n(\boldsymbol{\alpha})=\Tr(\prod_{j=1}^n\rho_Ae^{i\alpha_{jj+1} Q_A}),
\end{equation}
where $\boldsymbol{\alpha}=\{\alpha_1, \dots,\alpha_n\}$, $\alpha_{jj+1}=\alpha_j-\alpha_{j+1}$ and $\alpha_{n+1}\equiv\alpha_1$.
The latter can be computed adapting the standard methods employed to obtain the $n$-R\'enyi entropy in random unitary circuits~\cite{fisher2023random,potter2022entanglement}.  In essence, $U$ is associated with a tensor with two pairs of indices, one for each regions $A$ and $B$. The tensor can be represented graphically by a rectangle with lower (upper) legs associated to their input (output) degrees of freedom. The state $U\ket{0}$ is expressed using a single copy of such a tensor, while $U\ket{0}\bra{0}U^{\dagger}$ involves the bra and the ket, so it can be represented by two layers of $U$. Therefore, $Z_n(\boldsymbol{\alpha})$ is a sequence of $2n$ layers sewn together as shown in Fig.~\ref{fig:circuit} for $n=2$. 

In formulas, the average of $Z_n(\boldsymbol{\alpha})$ amounts to compute
\begin{equation}\label{eq:averaged_charged_mom}
\mathbb{E}[Z_n(\boldsymbol{\alpha})]=\bra{-+;\boldsymbol{\alpha}}\mathbb{E}[U^{\otimes n}\otimes (U^*)^{\otimes n}]\ket{0}^{\otimes 2n},
\end{equation}
where we have introduced the states
\begin{equation}\label{eq:pm_alpha}
\ket{-+;\boldsymbol{\alpha}}=\bigotimes_{k\in A}\ket{-;\boldsymbol{\alpha}}_k\bigotimes_{k\in B}\ket{+}_k,
\end{equation}
with
\begin{equation}\label{eq:plus}
\ket{+}_k=\sum_{\{a_{j}=0\}}^1\bigotimes_{j=1}^n(\ket{a_j}_k\otimes \ket{a_j}_k)
\end{equation}
and, assuming $a_{n+1}\equiv a_1$,
\begin{equation}\label{eq:minus}
\ket{-;\boldsymbol{\alpha}}_k=\sum_{\{a_j=0\}}^1
\bigotimes_{j=1}^n(e^{i\alpha_{jj+1}a_j}\ket{a_j}_k\otimes e^{i\alpha_{j+1j+2}a_{j+1}}\ket{a_{j+1}}_k).
\end{equation}
The definition of the conjugate operator $U^\ast$ depends on the choice of the local basis, which is fixed by Eqs.~\eqref{eq:plus} and ~\eqref{eq:minus}.

The Haar average of the tensor product of unitary matrices in Eq.~\eqref{eq:averaged_charged_mom} can be calculated with the well-known formula~\cite{weingarten-78, cs-06}
\begin{equation}\label{eq:Haar_average}
\mathbb{E}[U^{\otimes n}\otimes (U^*)^{\otimes n}]=
\sum_{\sigma_1, \sigma_2\in \mathcal{S}_n}{\rm Wg}(\sigma_1\sigma_2^{-1})\ket{\sigma_1}\bra{\sigma_2},
\end{equation}
where $\mathcal{S}_n$ is the symmetric group, $\ket{\sigma}=\otimes_{k=1}^L\ket{\sigma}_k$, and
\begin{equation}
\ket{\sigma}_k=\sum_{\{a_j=0\}}^1\bigotimes_{j=1}^n(\ket{a_j}_k\otimes \ket{a_{\sigma(j)}}_k).
\end{equation}
The Weingarten coefficients ${\rm Wg}(\sigma)$ in Eq.~\eqref{eq:Haar_average} can be determined analytically.
Up to exponentially small corrections in $L$, 
they are ${\rm Wg}(\sigma)=2^{-nL}\delta_{\sigma, {\rm Id}}$ where ${\rm Id}\in\mathcal{S}_n$ stands for the identity permutation~\cite{jyvl-20}. 
This significantly simplifies the analytic calculation of $\mathbb{E}[Z_n(\boldsymbol{\alpha})]$, as we only have to consider the diagonal terms $\sigma_1=\sigma_2$ in Eq.~\eqref{eq:Haar_average}. Being the corrections exponential, the asymptotic form works perfectly
even for small $L$, as we numerically check for several values of $n$.

\begin{figure}
     \centering
     \includegraphics[width=\linewidth]{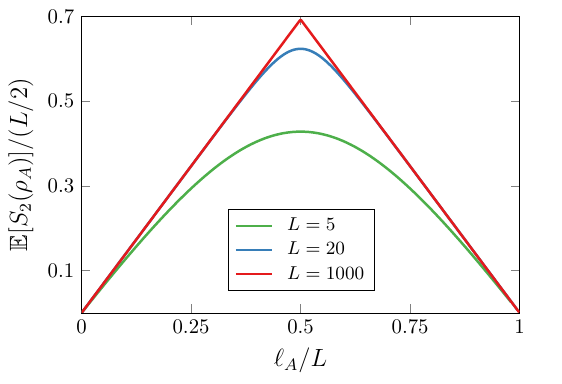}
     \includegraphics[width=\linewidth]{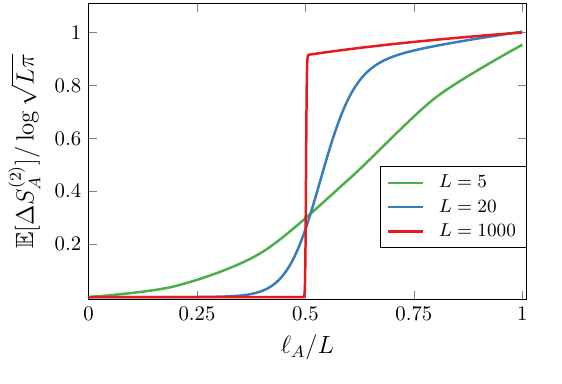}
     \caption{Averaged 2-R\'enyi entanglement entropy (upper panel) and entanglement asymmetry of $A$ (lower panel) for Haar random pure states as a function of the subsystem size $\ell_A$ for different fixed total system sizes $L$. In the upper panel, $\mathbb{E}[S_2(\rho_A)]$ is obtained from Eq.~\eqref{eq:averaged_charged_mom_2} taking $n=2$ and $\alpha=0$. In the lower panel, the curves represent Eq.~\eqref{eq:final_av_asymm}.}
     \label{fig:av_ent_asymm}
\end{figure}

Combining all the previous ingredients and computing the overlaps involving the states in Eq.~\eqref{eq:pm_alpha} and \eqref{eq:Haar_average}, we find
\begin{multline}\label{eq:averaged_charged_mom_2}
\mathbb{E}[Z_n(\boldsymbol{\alpha})]=\frac{1}{2^{nL}}
\sum_{\sigma\in \mathcal{S}_n}\left(\sum_{\{a_j=0\}}^1\prod_{j=1}^n\delta_{a_j,a_{\sigma(j)}}\right)^{L-\ell_A}\\
\times \left(\sum_{\{a_j=0\}}^1\prod_{j=1}^n e^{i2\alpha_{jj+1}a_j}
\delta_{a_{j+1}, a_{\sigma(j)}}\right)^{\ell_A}.
\end{multline}
Note that, taking $\alpha=0$ in this formula, we straightforwardly obtain an analytic expression for the \textit{averaged} $n$-R\'enyi entanglement entropy $S_n(\rho_A)$, which we plot in the upper panel of Fig.~\ref{fig:av_ent_asymm}. For large $L$, it tends to the original result by Page in Eq.~\eqref{eq:page_curve} (we recall that the Page curve is valid for R\'enyi entropies of arbitrary order, while exact analytic expressions at finite size exist for $n=1$, see Ref.~\cite{bianchi2022volume} for a recent review).

Plugging Eq.~\eqref{eq:averaged_charged_mom_2} in Eq.~\eqref{eq:renyi_rhoAQ}, we  obtain the averaged
asymmetry. In particular, for $n=2$ the resulting expression
is very simple,
\begin{equation}\label{eq:final_av_asymm}
\mathbb{E}[\Delta S_A^{(2)}]=-\log\left[\frac{1}{2^{2\ell_A-L}+1}\left(1+2^{-L}\frac{(2\ell_A)!}{(\ell_A!)^2}\right)\right].
\end{equation}
We plot this expression in the lower panel of Fig.~\ref{fig:av_ent_asymm} as a function of the subsystem size $\ell_A$, choosing different fixed values of $L$.

\textit{Large-$L$ asymptotics for generic $n$ ---}
In the large-$L$ limit, the two dominant terms in Eq.~\eqref{eq:averaged_charged_mom_2} correspond
to the permutations $\sigma(j)=j$ (identity) and $\sigma(j)=j+1$, and the resulting asymptotic charged moments are 
\begin{equation}\label{eq:averaged_charged_mom_large_L}
\mathbb{E}[Z_n(\boldsymbol{\alpha})]\sim 2^{(1-n)\ell_A}+2^{(1-n)(L-\ell_A)}\prod_{j=1}^n\cos(\alpha_{jj+1})^{\ell_A}.
\end{equation}
Plugging this expression into Eq.~\eqref{eq:renyi_rhoAQ}, we find that
\begin{equation}
\label{eq:final_av_asymm_large_L}
\mathbb{E}[\Delta S_A^{(n)}]\sim
\left\{
\begin{array}{ll}0, &\quad \ell_A<L/2,\\
1/2\log(\ell_A\pi n^{1/(n-1)}/2), &\quad \ell_A>L/2.
\end{array}\right.
\end{equation}
From this equation, the analytic continuation for the asymptotics of the von Neumann asymmetry (i.e., $n\to1$) yields $\mathbb{E}[\Delta S_A^{(1)}]=1/2\log(\ell_A\pi /2)+1/2$ for $\ell_A>L/2$. 
For $n=2$, this result  could have been alternatively derived using the methods of Ref.~\cite{lntt-22}.

\begin{figure}
\includegraphics[width=\linewidth]{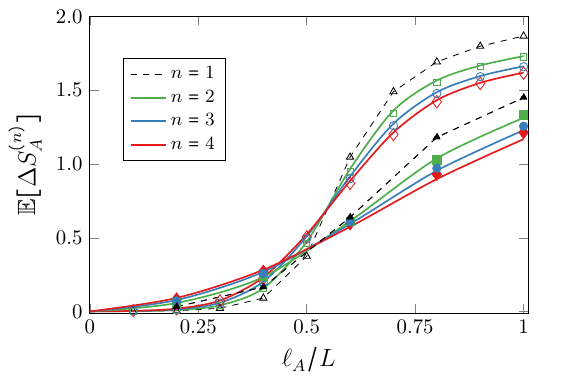}
     \caption{Averaged $n$-R\'enyi entanglement asymmetry of $A$ for Haar random pure states as a function of the subsystem size $\ell_A$ for fixed total system size $L=5$ (filled symbols), $L=10$ (empty symbols) and different $n$. The solid continuous lines are obtained plugging Eq.~\eqref{eq:averaged_charged_mom_2} into Eq.~\eqref{eq:renyi_rhoAQ}. The symbols are the numerical data obtained via  numerically-exact computations. For $n=1$, the dashed line only joins the numerical data as a guide for the eyes.  Statistical errors are not visible at the scales of the plot.}
     \label{fig:av_ent_asymm2}
\end{figure}

\textit{Numerical check ---}  We have checked our prediction in Eq.~\eqref{eq:averaged_charged_mom_2} against numerically-exact computations at finite size. We generate the initial state $U\ket{0}$, where $U$ is drawn from the uniform distribution over the unitary group $U(2^L)$.  For each $\ell_A$, we take a number of samples making statistical errors sufficiently small (in practice, $\sim 10^2$ is enough for our purposes). The average asymmetry $\mathbb{E}[\Delta S_A^{(n)}]$ is then obtained by averaging over all the samples.
In Fig. \ref{fig:av_ent_asymm2}, we show the results for $L=5, 10$ and different values of $n$ (symbols) and we compare them using the analytical prediction in Eq.~\eqref{eq:averaged_charged_mom_2}, finding a remarkable agreement even for small system sizes. Such agreement confirms that all neglected terms are exponentially suppressed in $L$.
We omit the error associated with the finite-number of samples since it is smaller than the symbol size.

\textit{Discussion and physical interpretation} ---
Eq. \eqref{eq:final_av_asymm_large_L} is the main result of this work: it tells us that, for $\ell_A<L/2$, the asymmetry vanishes, meaning that the $U(1)$ symmetry is typically restored in the subsystem $A$. 
At the Page time, $\ell_A=L/2$, the asymmetry shows a sharp jump, i.e. the $U(1)$ symmetry is suddenly broken,
and then it increases logarithmically for $\ell_A>L/2$. 
A remarkable consequence of Page's result~\eqref{eq:page_curve}
is that typical states have almost maximal bipartite entanglement. 
Something similar happens for the entanglement asymmetry:
In Ref.~\cite{amc-23}, it was shown that the entanglement asymmetry for the product states that maximally break a $U(1)$ symmetry is exactly $1/2\log(\ell_A \pi n^{1/(n-1)}/2)$ for large $\ell_A$ (there are no indications that this bound is valid for entangled states, but some arguments have been provided for physical states~\cite{cv-23}). 
According to Eq.~\eqref{eq:final_av_asymm_large_L}, this is precisely the asymmetry
that typical states display for $\ell_A>L/2$ and large $L$. 
This implies that at the Page time the asymmetry displays the largest possible jump.

It may initially seem surprising that a typical random non-symmetric state possesses a symmetry. Yet,  this fact is intuitive in light of the decoupling inequality~\cite{hp-07}, which states 
\begin{equation}\label{eq:decoupling}
\mathbb{E}\left[\Big|\Big|\rho_A-\frac{\openone}{2^{\ell_A}}\Big|\Big|_1\right]^2\leq2^{\ell_A-\ell_B},
\end{equation}
where $||\cdot||_1$ stands for the $L_1$ norm.
Physically, this inequality implies that for large sizes the reduced density matrix $\rho_A$ is exponentially close to the (normalized) identity for $\ell_A<\ell_B$.
As a consequence, one expects that also their asymmetries are close (the asymmetry of the identity being vanishing because there are no off-diagonal terms).
Conversely, for $\ell_A>\ell_B$, the right hand side of the above inequality becomes exponentially large and there is no bound.

The sharp transition in the radiation from a symmetric to a non-symmetric state has  been noted in Ref.~\cite{islands2020} in the context of a specific gravity theory, using a quantity akin to $Z_2(\alpha)$ and the replica wormhole formalism. The jump at the Page time is associated with the emergence of an ``island''~\cite{aemm-19}, a region inside the black hole that contributes to the entanglement entropy of the radiation, see also~\cite{hs-21} and~\cite{hiy-21}.

{\it Outlooks ---} We have studied the averaged entanglement asymmetry in Haar random states. 
Our main result is given by Eq.~\eqref{eq:final_av_asymm_large_L} showing that in the thermodynamic limit the entanglement asymmetry is vanishing for $\ell_A<L/2$ and, after a finite jump at $\ell=L/2$, it grows logarithmically in $\ell_A$ for $\ell_A>L/2$. 
We note that our choice of a $U(1)$ symmetry generator $Q$ is completely arbitrary, since we do not have a Hamiltonian fixing the conserved charges. However, given the inequality \eqref{eq:decoupling}, it is natural to expect that a similar behavior is found for larger symmetry groups. This is especially true in the case of Abelian symmetries, while technical complications and non-trivial differences may appear in the case of non-Abelian symmetries. Therefore, by modelling an evaporating black hole with a pure random state, symmetries should be broken as we predict, for any choice of the charge.

Our result has interesting implications for an evaporating black hole: the emitted radiation is symmetric before the Page time, even in the absence of any symmetry;  conversely it undergoes a significant (potentially maximal) symmetry breaking shortly afterward. 
The transition between the two regimes is discontinuous
and it is reminiscent of the decoding transition by Hayden and Preskill~\cite{hp-07}.  If we choose a proper quantum field theory with a specific Hamiltonian for describing the black hole, our findings imply that it would be impossible to determine whether the theory has a conserved charge without collecting at least half of the black hole's radiation.
%
Note that the asymmetry of the black hole (obtained by looking at $\ell_B$) displays the opposite behavior: it is maximal before the Page time and it vanishes after. 

Looking forward, it is desirable to compute the asymmetry of the radiation in more realistic settings, including dynamical models of evaporating black holes (e.g., those in Refs.~\cite{hp-07,psq-20}) or by leveraging the holographic correspondence, akin to the approaches taken for the Page curve, see e.g.~\cite{ams-20, gk-20, ammz-20,rsv-20,cfh-20}.
Additionally, it is worthwhile to explore methods for experimentally investigating the entanglement asymmetry and its transition at the Page time in analog gravity settings
\cite{ltt-19,kgm-21,ltt-21}.

{\it Acknowledgements ---}
We thank Ahmed Almheiri, Souvik Banerjee,  Stefano Liberati, Shota Komatsu, Kyriakos Papadodimas, and John Preskill for fruitful discussions. 
PC and FA acknowledge support from ERC under Consolidator Grant number
771536 (NEMO).
SM thanks the support from the Caltech Institute for Quantum Information and Matter and the Walter Burke Institute for Theoretical Physics at Caltech. 

\emph{Note added.} After the submission of this manuscript, a complementary work appeared in the arXiv~\cite{liu2024symmetry} showing the validity of our results even in the context of random unitary circuits.


\begin{thebibliography}{100}

\bibitem{hawking-75}
S. W. Hawking, 
{\it Particle Creation by Black Holes}, 
\href{https://doi.org/10.1007/BF02345020}{Commun. Math. Phys. {\bf 43}, 199 (1975)}. 

\bibitem{hawking-76}
 S. W. Hawking, 
 {\it Breakdown of Predictability in Gravitational Collapse},
 \href{https://doi.org/10.1103/PhysRevD.14.2460}{Phys. Rev. D {\bf 14}, 2460 (1976)}.

 \bibitem{page-93}
 D. N. Page,
 {\it Average entropy of a subsystem},
\href{https://doi.org/10.1103/PhysRevLett.71.1291}{Phys. Rev. Lett. {\bf 71}, 1291 (1993)}.

 \bibitem{page-93-2}
 D. N. Page,
 {\it Information in black hole radiation},
\href{https://doi.org/10.1103/PhysRevLett.71.3743}{Phys. Rev. Lett. {\bf 71}, 3743 (1993)}.

\bibitem{aemm-19}
A. Almheiri, N. Engelhardt, D. Marolf, and H. Maxfield, 
{\it The entropy of bulk quantum fields and the entanglement wedge of an evaporating black hole},
\href{https://doi.org/10.1007/JHEP12(2019)063}{JHEP {\bf 12} (2019) 06}.

\bibitem{penington-20}
G. Penington,
{\it Entanglement wedge reconstruction and the information paradox},
\href{https://doi.org/10.1007/JHEP09(2020)002}{JHEP {\bf 09} (2020) 002}.

\bibitem{ahmst-20}
A. Almheiri, T. Hartman, J. Maldacena, E. Shaghoulian,
and A. Tajdini,
{\it Replica wormholes and the entropy of Hawking
radiation},
\href{https://doi.org/10.1007/JHEP05(2020)013}{JHEP {\bf 05} (2020) 013}.



\bibitem{pssz-22}
G. Penington, S. H. Shenker, D. Stanford, and Z. Yang,
{\it Replica wormholes and the black hole interior},
\href{https://doi.org/10.1007/JHEP03(2022)205}{JHEP {\bf 03} (2022) 205}.

\bibitem{mw-57}
C. W. Misner and J. A. Wheeler, 
{\it Classical physics as geometry: gravitation,
electromagnetism, unquantized charge, and mass as properties of curved empty space},
\href{https://doi.org/10.1016/0003-4916(57)90049-0}{Ann. Phys. {\bf 2}, 525 (1957)}.

\bibitem{bd-88}
T. Banks and L. J. Dixon, 
{\it Constraints on string vacua with space-time supersymmetry},
\href{https://doi.org/10.1016/0550-3213(88)90523-8}{Nucl. Phys. B {\bf 307}, 93 (1988)}.

\bibitem{klls-95}
R. Kallosh, A. Linde, D. Linde, and L. Susskind, 
{\it Gravity and global symmetries}, 
\href{https://doi.org/10.1103/PhysRevD.52.912}{Phys. Rev. D {\bf 52}, 912 (1995)}.

\bibitem{bs-11}
T. Banks and N. Seiberg, 
{\it Symmetries and strings in field theory and gravity}, 
\href{https://doi.org/10.1103/PhysRevD.83.084019}{Phys. Rev. D {\bf 83}, 084019 (2011)}.

\bibitem{ho-19}
D. Harlow and H. Ooguri,
{\it Constraints on symmetry from holography},
\href{https://doi.org/10.1103/PhysRevLett.122.191601}{Phys. Rev. Lett. {\bf 122}, 191601 (2019)}.

\bibitem{ho-21}
D. Harlow and H. Ooguri,
{\it Symmetries in quantum field theory and quantum gravity},
\href{ https://doi.org/10.1007/s00220-021-04040-y}{Comm. Math. Phys. {\bf 383}, 1669 (2021)}.

\bibitem{hs-21}
D. Harlow and E. Shaghoulian,
{\it Global symmetry, Euclidean gravity, and the black
hole information problem},
\href{https://doi.org/10.1007/JHEP04(2021)175}{JHEP {\bf 04} (2021) 175}.

\bibitem{amc-23}
F. Ares, S. Murciano, and P. Calabrese,
{\it Entanglement asymmetry as a probe of symmetry breaking},
\href{https://doi.org/10.1038/s41467-023-37747-8}{Nature Commun. {\bf 14}, 2036 (2023)}.

\bibitem{amvc-23}
F. Ares, S. Murciano, E. Vernier, and P. Calabrese, {\it Lack of symmetry restoration after a quantum quench: an entanglement asymmetry study},
\href{https://doi.org/10.21468/SciPostPhys.15.3.089}{SciPost Phys. {\bf 15},  089 (2023)}.

\bibitem{bkccr-23}
B. Bertini, K. Klobas, M. Collura, P. Calabrese, and C. Rylands,
{\it Dynamics of charge fluctuations from asymmetric initial states},
\href{https://doi.org/10.1103/PhysRevB.109.184312}{Phys. Rev. B {\bf 109}, 184312 (2024)}.

\bibitem{fac-23b}
F. Ferro, F. Ares, and P. Calabrese, {\it Non-equilibrium entanglement asymmetry for discrete groups: the example of the XY spin chain},
\href{https://doi.org/10.1088/1742-5468/ad138f}{J. Stat. Mech. (2024) 023101}.

\bibitem{cm-23}
L. Capizzi and M. Mazzoni, {\it Entanglement asymmetry in the ordered phase of many-body systems: the Ising Field Theory},
\href{https://doi.org/10.1007/JHEP12(2023)144}{JHEP {\bf 12} (2023) 144}.

\bibitem{rka-23}
C. Rylands, K. Klobas, F. Ares, P. Calabrese, S. Murciano, and B. Bertini, {\it Microscopic origin of the quantum Mpemba effect in integrable systems},
\href{https://doi.org/10.1103/PhysRevLett.133.010401}{Phys. Rev. Lett. {\bf 133}, 010401 (2024)}.

\bibitem{cv-23}
L. Capizzi and V. Vitale,
{\it A universal formula for the entanglement asymmetry of matrix product states},
\href{https://arxiv.org/pdf/2310.01962.pdf}{ArXiv:2310.01962}.

\bibitem{makc-23}
S. Murciano, F. Ares, I. Klich, and P. Calabrese, {\it Entanglement asymmetry and quantum Mpemba effect in the XY spin chain},
\href{https://doi.org/10.1088/1742-5468/ad17b4}{J. Stat. Mech. (2024) 013103}.

\bibitem{cc-23}
M. Chen, H.-H. Chen,
{\it R\'nyi entanglement asymmetry in 1+1-dimensional conformal field theories},
\href{https://arxiv.org/abs/2310.15480}{Phys. Rev. D {\bf 109}, 065009 (2024)}.


\bibitem{lntt-22}
P. H. C. Lau, T. Noumi, Y. Takii and K. Tamaoka, {\it Page curve and symmetries}, 
\href{https://arxiv.org/pdf/2206.09633.pdf}{JHEP {\bf 10} (2022) 015}.


\bibitem{bd-19}
E. Bianchi and P. Dona, {\it Typical entanglement entropy in the presence of a center: Page
curve and its variance},
\href{https://doi.org/10.1103/PhysRevD.100.105010}{Phys. Rev. D {\bf 100},  105010 (2019)}.

\bibitem{mcp-22}
S. Murciano, P. Calabrese, and L. Piroli, {\it Symmetry-resolved Page curves}, 
\href{https://doi.org/10.1103/PhysRevD.106.046015}{Phys. Rev. D {\bf 106}, 046015 (2022)}.

\bibitem{ll-23}
P. Li and Y. Ling,
{\it Refined symmetry-resolved Page curve and charged black holes}, \href{https://doi.org/10.1088/1674-1137/ad2e83}{Chinese Phys. C {\bf 48} 053109 (2024)}.

\bibitem{hms-23}
C. Han, Y. Meir, and E. Sela, 
{\it Realistic Protocol to Measure Entanglement at Finite Temperatures},
\href{https://doi.org/10.1103/PhysRevLett.130.136201}{Phys. Rev. Lett. {\bf 130}, 136201 (2023)}.


\bibitem{r1}
J. A. Vaccaro, F. Anselmi, H. M. Wiseman, and K. Jacobs, {\it Tradeoff between extractable mechanical work, accessible entanglement, and ability to act as a reference system, under arbitrary superselection rules},
\href{https://journals.aps.org/pra/abstract/10.1103/PhysRevA.77.032114}{Phys. Rev. A {\bf 77}, 032114 (2008)}.

\bibitem{r2}
G. Gour, I. Marvian, and R. W. Spekkens
{\it Measuring the quality of a quantum reference frame: the relative entropy of frameness},
\href{https://doi.org/10.1103/PhysRevA.80.012307}{Phys. Rev. A {\bf 80}, 012307 (2009)}

\bibitem{r3}
I. Marvian and R. W Spekkens, 
{\it Extending Noether’s theorem by quantifying the asymmetry of quantum states},
\href{https://www.nature.com/articles/ncomms4821}{Nat. Comm. {\bf 5}, 3821 (2014)}

\bibitem{r4}
R. Takagi,
{\it Skew informations from an operational view via resource theory of asymmetry},
\href{ttps://doi.org/10.1038/s41598-019-50279-w}{Scien. Rep. 9, 14562 (2019)}.

\bibitem{fisher2023random}
M. P. A. Fisher, V. Khemani, A. Nahum, and S. Vijay, {\it Random Quantum Circuits}, \href{https://www.annualreviews.org/doi/10.1146/annurev-conmatphys-031720-030658}{Ann. Rev. Cond. Matt. Phys. {\bf 14}, 335 (2023)}

\bibitem{potter2022entanglement}
A. C. Potter and R. Vasseur, {\it Entanglement Dynamics in Hybrid Quantum Circuits}, in Entanglement in Spin Chains: From Theory to Quantum Technology Applications, edited by A. Bayat, S. Bose, and H. Johannesson (Springer International Publishing, Cham, 2022), pp. 211–249.





\bibitem{weingarten-78}
D. Weingarten, 
{\it Asymptotic behavior of group integrals in the limit of infinite rank}, 
\href{https://doi.org/10.1063/1.523807}{J. Math. Phys. {\bf 19}, 999 (1978)}.

\bibitem{cs-06}
B. Collins and P. Sniady,
{\it Integration with Respect to the Haar Measure on Unitary, Orthogonal and Symplectic Group}, 
\href{https://doi.org/10.1007/s00220-006-1554-3}{Comm. Math. Phys. {\bf 264}, 773 (2006)}.




\bibitem{jyvl-20}
C.-M. Jian, Y.-Z. You, R. Vasseur, and A. W. W. Ludwig,
{\it Measurement-induced criticality in random quantum circuits},
\href{https://doi.org/10.1103/PhysRevB.101.104302}{Phys. Rev. B {\bf 101}, 104302 (2020)}.

\bibitem{bianchi2022volume}
E. Bianchi, L. Hackl, M. Kieburg, M. Rigol, and L. Vidmar, {\it Volume-Law Entanglement Entropy of Typical Pure Quantum States}, \href{https://journals.aps.org/prxquantum/abstract/10.1103/PRXQuantum.3.030201}{PRX Quantum {\bf 3}, 030201 (2022)}.

\bibitem{hp-07}
P. Hayden and J. Preskill, 
{\it Black holes as mirrors: quantum information in random subsystems},
\href{https://doi.org/10.1088/1126-6708/2007/09/120}{JHEP {\bf 09} (2007) 120}.

\bibitem{islands2020}
Y. Chen and H. W. Lin, 
{\it Signatures of global symmetry violation in relative entropies and replica wormholes,} 
\href{https://link.springer.com/article/10.1007/JHEP03(2021)040}{JHEP {\bf 03} (2021) 040.}

\bibitem{hiy-21}
P.-S. Hsin, L. V. Iliesiu, and Z. Yang,
{\it A violation of global symmetries from replica wormholes and the fate of black hole remnants},
\href{https://doi.org/10.1088/1361-6382/ac2134}{Class. Quantum Grav. {\bf 38}, 194004 (2021)}.


\bibitem{psq-20}
L. Piroli, C. S\"underhauf, and X.-L. Qi, 
{\it A Random Unitary Circuit Model for Black Hole Evaporation}, 
\href{https://doi.org/10.1007/JHEP04%282020%29063}{JHEP {\bf 04} (2020) 63}.

\bibitem{ams-20}
A. Almheiri, R. Mahajan, and J. E. Santos,
{\it Entanglement islands in higher dimensions},
\href{https://doi.org/10.21468/SciPostPhys.9.1.001}{SciPost Phys. {\bf 9}, 001 (2020)}.

\bibitem{gk-20}
H. Geng and A. Karch,
{\it Massive islands},
\href{https://doi.org/10.1007/JHEP09%282020%29121}{JHEP {\bf 09} (2020) 121}.

\bibitem{ammz-20}
A. Almheiri, R. Mahajan, J. Maldacena, and Y. Zhao, {\it The Page curve of Hawking radiation from semiclassical geometry}, \href{https://doi.org/10.1007/JHEP03%282020%29149}{JHEP {\bf 03} (2020) 149}.

\bibitem{rsv-20}
M. Rozali J. Sully, M. Van Raamsdonk, C. Waddell, and D. Wakeham, {\it Information radiation in BCFT models of black holes}, 
\href{https://doi.org/10.1007/JHEP05(2020)004}{JHEP {\bf 05} (2020) 004}.

\bibitem{cfh-20}
H.Z. Chen, Z. Fisher, J. Hernandez, R. C. Myers, and S.-M. Ruan, 
{\it Information flow in black hole evaporation}, 
\href{https://doi.org/10.1007/JHEP03%282020%29152}{JHEP {\bf 03} (2020) 152}.

\bibitem{ltt-19}
S. Liberati, G. Tricella, and A. Trombettoni, 
{\it The Information Loss Problem: An Analogue Gravity Perspective},
\href{https://doi.org/10.3390/e21100940}{Entropy  {\bf 21}, 940 (2019)}.

\bibitem{kgm-21}
V. I. Kolobov, K. Golubkov, J. R. Munoz de Nova, and J. Steinhauer,
{\it Spontaneous Hawking radiation and beyond: Observing the time evolution of an analogue black hole},
\href{https://doi.org/10.1038/s41567-020-01076-0}{Nature Phys. {\bf 17}, 362 (2021)}.

\bibitem{ltt-21}
S. Liberati, G. Tricella, and A. Trombettoni, 
{\it
Back-reaction in canonical analogue black holes},
\href{https://doi.org/10.3390/app10248868}{Appl. Sciences {\bf 10}, 8868 (2020)}.

\bibitem{liu2024symmetry}
S. Liu, H.-K. Zhang, S. Yin, and S.-X. Zhang, 
{\it Symmetry restoration and quantum Mpemba effect in symmetric random circuits},
\href{https://doi.org/10.48550/arXiv.2403.08459}
{arXiv:2403.08459}.

\end{thebibliography}
\end{document}